# Symmetrical laws of structure of helicoidally-like biopolymers in the framework of algebraic topology. I. Root lattice $E_8$ and the closed sequence of algebraic polytopes

M.I.Samoylovich, A.L.Talis

*Central Research and Technology Institute "TechnoMash", Moscow*
*E-mail: samoylovich@technomash.ru*
*Institute of Organoelement Compounds of Russian Academy of Sciences, Moscow*

*In the framework of algebraic topology the closed sequence of 4-dimensional polyhedra (algebraic polytopes) was defined. These polytopes were determined by the second coordination sphere of 8-dimensional lattice $E_8$. The ordered non-crystalline structure is determined by a chain of constructions of algebraic topology: an algebraic polytope, a homogeneous manifold in a 3D Euclidean space $E^3$, locally-homogeneous manifold, locally minimal surface, 1-parameter family of helicoids, bundle (cover) with a base of cell complexes, local-lattice packing of cell complexes into a substructure of $E^3$, determined by helicoids. The formalism being developed allows one to surmount restrictions of classical crystallography and to single out a class of ordered non-crystalline structures, invariant with respect to structures determined by the lattice $E_8$. The topological stability of such substructures is determined by their relatedness to Weierstrass' representation, as well as the condition that the instability index of the surface equals zero. Formation of such structures corresponds to lifting a configuration degeneracy, and the stability of a state – to existence of a point of bifurcation.*

## 1 Introduction

Living matter in the condensed state is separated from the non-living (usually with crystalline and less frequently with glass-like ordering type) mostly by a special type of ordering, namely by transition from translational invariance to local periodicity (finiteness). Rod-like structures are also used, characterized by non-integral axes, a limited number of building blocks (as a rule, with helicoidally similar configurations) and with even lower number of structural topological rules of their construction and hierarchical ordering, characteristic of $E^3$

Klein put forward a principle according to which every transformation group can serve as an automorphism group and H. Weil has shown that in a more general sense groups of automorphisms are abstract groups (and not necessarily transformation groups), which characterize "geometry" (in Klein's sense), while the type of the variable in such geometry in characterized by its transformation law. Every abstract group allows an exact realization whose action field is the group's manifold itself. The 3D Euclidean vector space forms a non-commutative (simple) Lie algebra (with the respect to the operation of vector multiplication), isomorphic to the algebra of quaternions with the norm equal to unity, which is isomorphic to the group of rotations of $E^3$. In such algebra there are no other subalgebras, except 1-dimensional ones, and every 1D (2D real, taking into account complexity) linear subspace form a subalgebra.

The group G itself can be considered a transformation group of the manifold M if for nay element of the group there is a diffeomorphism of M of certain type. Given a transitive action of G on M, the manifold is then called a uniform space of the group. The fact that the helicoid is the unique minimal ruled surface leads not only to the uniqueness of constructions, but also makes it necessary to account for singularities in such systems. For them the symmetry elements characteristic for harmonic functions are substituted by a reflection principle, where the union of two minimal surfaces, symmetric with respect to segments of their boundaries is also a minimal surface. In addition to the reflection property, such surfaces have the following features. If the intersection of two smooth minimal surfaces $M_1 \cap M_2$ contains some open subset, then their union is also a smooth minimal surface. If $M_1$ are $M_2$ are both minimal ruled surfaces and have congruent frameworks (families of generatrices everywhere dense on the surface), they are congruent themselves [1,2].

In treatment that follows, symmetric spaces, in particular, locally symmetric (locally homogenous) manifolds are used. Symmetric spaces include spaces with homogeneous (for 1-connected ones) metrics (in general, such spaces where the covariant derivative of the curvature tensor of symmetric connection vanishes). For 1-connected spaces the definition of such manifolds is reduced to the requirement that for any point $x \in M$ there is an isometric mapping (motion) $s_x : M \to M$ with x as an isolated fixed point such that all tangent vectors undergo a reflection in it, so that under the induced (diffeomorphic) mapping $s_{x*}$ the vector $\alpha$ goes to $-\alpha$, the transformation $s_x$ is called "symmetry" in x.

A Non-1-connected manifold $M^*$ is obtained from a 1-connected manifold M by by a group of discrete motions Γ, which leads to a non-homogeneous factor M/Γ. The group thus obtained may consist of shifts combined with reflections (as in Klein bottle) and, generally speaking, does not commute with the entire group of motions of the manifold $M^*$. Such an approach determines the transformation of a discrete manifold as a homogeneous space into a family of subspaces which is a locally homogeneous manifold. Local homogeneity may be introduced for any Riemannian manifold, hence, under the aforementioned conditions, all transformations $s_x$ are motions for all points of M.

Thus, the factor M/Γ will not be homogeneous – such manifolds are termed locally-homogeneous (locally-symmetric). Such an approach determines conversion of a discrete manifold as a homogeneous spaces into a collection of subspaces, each of which is only a locally homogeneous manifold. Because for any Riemannian manifold one can introduce local homogeneity, then under the said conditions all transformations $s_x$ are motions for all points of M. For instance, in the simplest case of dimension n=2, when the curvature tensor is determined by a single constant (the latter makes it necessary to study locally or globally minimal surfaces).

Minimal surfaces (with zero Gaussian curvature) allow one to define conformal mappings on them so that points on such a surface correspond to zeros of the derivative function in Weierstrass's representation (see appendix A in part 2). An action of a group given by linear transformations on a vector space на is its linear representation. Hence the properties of a subset of points can be carried to a manifold by defining appropriate subsets in the tangent (or cotangent) spaces to those points (using diffeomorphisms and differential forms). Thus one may arrive at using algebraic varieties given, in particular, by adjoint representations of appropriate transformation groups [1-3].

Complete minimal surface form a one-parameter family of helicoids. Such constructions can be viewed as a collection (direct sum) of helicoidally similar systems represented by cylindrical (tube-like) surfaces. On each of them (for a certain sort of atoms) a curve can be drawn on a helicoid-type surface with fixed algebraic relations between the length of the turn and inter-turn distance.

The complete minimal surfaces of revolution form a 1-parameter family of catenoids, and helicoidal systems as minimal ruled surfaces, also form a one-parameter family [2]. In what follows we shall limit ourselves to considering topologically meaningful systems, which are characterized by a minimal surface determined by the theorem about tubular surface. They include local lattice packing of cell complexes, bounded by the surface whose singular points are related by transformations given by a special homogeneous manifold. Such a manifold, an "algebraic" polytope generated by a subsystem of the root lattice $E_8$ of maximal exceptional Lie algebra $e_8$, which allows one to define a homogeneous manifold, bounded in space as well as in number of elements [4-6]. Using the $E_8$ lattice is explained not only by its being an octonion lattice which is at the top of the row of possible "numbers": real – complex – quaternions – octonions [5], but also by use of cell constructions.

The developed approach allows one to construct a family of minimal helicoidally similar surfaces, which can be considered a union of similar surfaces with similar topological characteristics. The collection of degenerate critical points of such system (which corresponds to a certain number of invariants $E_8$) by small perturbations can be turned into a set of non-degenerate critical points (corresponding to another collection of invariants $E_8$). Therefore, from using calculations of the system's minimum for a problem with finite displacements of atoms one may go to constructing systems corresponding to minima as bifurcation points due to small bifurcations which lift configuration degeneracies and correspond to local topological properties of the lattice $E_8$ as a symmetric prophase.

Using the first and the second coordination spheres of the root lattice $E_8$ normed on the unit sphere $S^7$ leads to the group SU(2), for which $S^7$ is the principal bundle space (see appendix B in part 2). The latter implies the possibility to introduce unitary periodicity (to be more precise, orthogonal periodicity) for unitary groups, which are important when using exponential representations like exp*ad*, for instance, when constructing simple Chevalley groups of given type, defined over an arbitrary field K [4].

In 1930 Coxeter [7] defined the polytope {240} – a diamond-like structure with 240 vertices on $S^3$. In present paper, which is a continuation of [8-15], it will be shown that the polytope {240} is the beginning of the closed sequence of algebraic polytopes, determined by the vectors of the 2$^{nd}$ coordination sphere of the lattice $E_8$. Summing the above, one may single out the following chain of constructions of algebraic topology: homogeneous manifold in $E^3 \rightarrow$ algebraic polytope from the sequence of polytopes within the 2$^{nd}$ coordination sphere of the lattice $E_8 \rightarrow$ locally-homogeneous manifold $\rightarrow$ locally-minimal surface $\rightarrow$ one-parameter family of helicoids $\rightarrow$ bundle (cover) with base of cell complexes $\rightarrow$ local lattice packing of cell complexes. Existence of such a chain implies a possibility to realize in $E^3$ a special class of structures given by local lattice packing of cell complexes, satisfying symmetries determined by the $E_8$ lattice.

## 2. The algebraic polytopes and rod substructures in $E^3$.

Each of the 240 vectors of the first coordination sphere of the system $E_8$, which determine the vertices of the eight dimensional Gosset polytope, can be placed in correspondence with a pair of quaternions specifying the octave [5, 11, 13]. This makes it possible to divide 240 vectors into ten 24-vector subsets $P_1, \ldots, P_{10}$ (Fig. 1); that is,

$$P_1=(T_2,0), \quad P_2=(0,T_2); \quad P_{3-6}=(T_1,\pm(1,i)T_1) \quad P_{7-10}=(T_1, \pm j(1,i)T_1) \quad (1),$$

where $i^2=j^2=k^2=-1$, $ji = k$, $T_1=1/2\{\pm 1, \pm i, \pm j, \pm k; 1/2(\pm 1, \pm i, \pm j, \pm k)\}$ is the Hurwitz group of unit quaternions; and $T_2=(1 + i)T_1$. Each subset $P_n$ corresponds to the four dimensional polytope {3, 4, 3} with the center located at one of the ten vertices of the five dimensional analogue of the the {3, 3, 3, 4} polytope (octahedron). This correspondence is determined by the Hopf fibration $S^7 \rightarrow S^4$ (fiber $S^3$), which is the principal fiber bundle for the group SU(2) and the associated fiber bundle for the group SO(4).

The tangent vectors to a one-parameter subgroup of the Lie group G of motions of a symmetric manifold M lie in a linear subspace L of the algebra g generated by all tangent vectors to geodesic lines emanating from an isolated fixed point $x_0$. The subalgebra $g_0$ corresponds to a stationary subgroup $H(x_0)$ of the group G, so that $g=g_0+g_1$, where $g_1$ is a subalgebra of g. Here the linear operator $\sigma: g \rightarrow g$ equal to 1 on $g_0$ and -1 on $g_1$ is an automorphism and preserves commutativity. For $\sigma^2=1$ (involution) the existence of such as automorphism leads to $Z_2$ – grading of the algebra g. For these groups of motions there is a subgroup of index 2 ($Z^2$), which determines the torus $T^2$, which is, for example, a 2-sheet cover of the Klein bottle ($RP^2$), because there are two orbits for $Z^2$.

Recall that any conformal mapping, close to unity on $E^n$ or $S^n$ can be represented as exp (tA), so that vector fields describing in A pseudo-rotations, translations, dilatation and inversion, form a Lie algebra for these transformation groups. Looking at vector fields given by motions of the Euclidean metric with vanishing "deformation tensor" (isometric), one can go beyond local transformations using, for example, the stereographic projection to move from $S^n$ to $E^n$. It can be shown that linear vector fields corresponding to orthogonal transformations are tangent to a sphere centered at the origin.

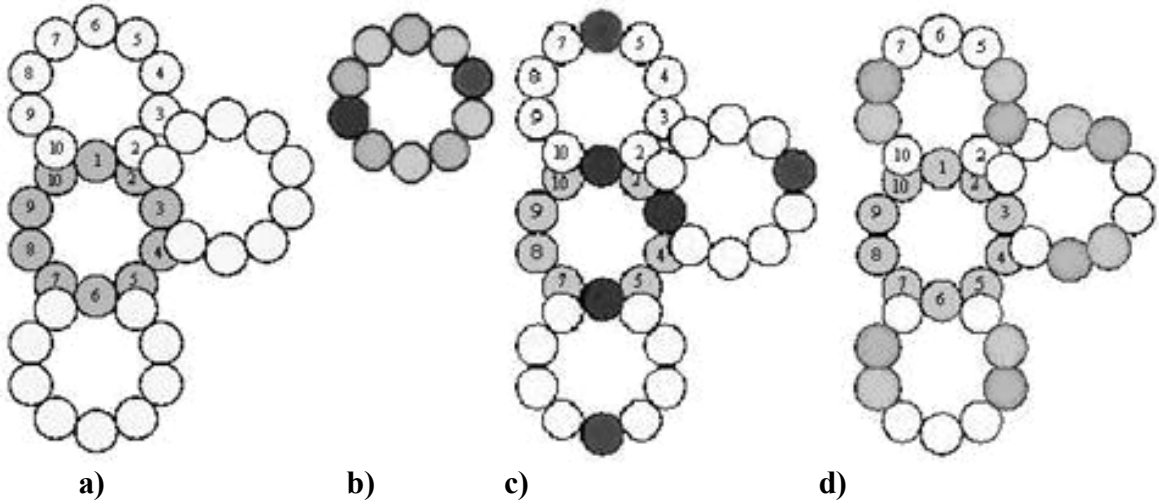

**a)          b)          c)          d)**

*Fig.1 a) A scheme of a quaternion representation of the 1$^{st}$ and 2$^{nd}$ coordination spheres of the lattice $E_8$. The first coordination sphere of the lattice $E_8$ is shown as a union of 10 gray circles, each of which corresponds to a 24-element subsystem of $E_8$. The second coordination sphere is represented as a union of 10 congruent first coordination spheres (The figure shows just 3 of them).*
*b) A schematic representation of Gossett's construction [6]. The black and the dark dots relate to elements put into correspondence with the polytopes {3, 4, 3} and {3, 4, 3}\*, dark-gray and light-gray ones with sn-{3, 4, 3} and sn-{3, 4, 3}\*.*
*c) The union of all 10 black dots corresponds to the polytope {240}, dark – to a non-diamond-like polytope {240}\*, black and dark – to a diamond-like polytope {480}.*
*d) The union of all dark-gray and light-gray circles corresponds to the polytope {960}.*

The space $S^3$ without the circle $S_1$ and a point (or, what is the same, the space $E^3$ without $S^1$) is homotopic to the union $S^1 \cup S^2$ of a circle and a sphere. Universal covering above the union $S^1 \cup S^2$ can be written as a set of spheres $S_2$ attached to a straight line at integer points $\{x_i$ for $i =1,2, …,k\}$ (Fig. 2. a.). Any compact oriented surface embedded in $E_3$ is homeomorphic to a connected sum of tori; therefore, a transition from $S^3$ to a universal covering above $S^1 \cup S^2$, which is, as well as $S^3$, a singly connected surface, is possible only in the case of selection of tori of the corresponding algebras on $S_3$, $S_2$, and $S_1$ (as group varieties):

$$S^2 \text{ (layer } S^1) \leftarrow S^3 \supset (S^3 \backslash S^1) \backslash \{x_1\} = E^3 \backslash S^1 \sim S^1 \vee S^2 \leftarrow \sum_{i}^{k} S_i^2 \quad (2),$$

where $A \backslash B$ is a set $A$ without a subset $B$, $\sim$ is the symbol of homotopy equivalence, and is the map →symbol. [1]

In the discrete implementation (2), the points $x_i$ to which spheres $S^2$ are attached can be put into correspondence with $k_{js}$ from $I_s$ roots of the subsystem of $E_8$. Therefore, $I_s= k_{js}+k_{js}m_{js}$, where $k_{js}$ is the number of roots, which are considered as zero roots (determining immobile points), and $m_{js}$ is the exponent of the system $E_8$. The covering $\sum_{i=1}^{k} S_i \rightarrow S^1 \vee S^2$ consists of $k_{js}$ spheres; therefore, an integer number of points p=$8I_n/\gamma_1\gamma_2 k_{js}$ from $E_8$ will be mapped onto the equator of each of them. Map of a sphere onto the equator can be two-sheet; therefore, $\gamma_1$ and $\gamma_2$ are equal to 1 or 2 depending on whether the corresponding map is single- or two-sheeted. In this case, the points $8I_n/\gamma_1\gamma_2 k_{js}$ divide the equator of the sphere into equal parts, and the ($m_{js}$+1)th point selects the part of the equator that contains $m_{js}$ such parts. Owing to the introduction of the external metric, the equators of all spheres are equal, parallel, and equidistant; therefore, the expression:

$$L/d = \frac{8}{\gamma_1 \gamma_2} \cdot \frac{I_n}{I_s} \cdot \frac{m_{js}+1}{m_{js}} = \frac{8 I_n}{\gamma_1 k_{js}} \cdot \frac{1}{\gamma_2 m_{js}} \qquad (3)$$

determines the angle of rotation (360°/L)d of the helicoid axis (see appendix D in part 2). In [8] a special class of helicoids is considered, the Gosset helicoids, the screw axes of which are determined by the relation (3). This relation sets the map (retaining the surface local minimality) "on the cylinder" for an algebraic polytope, which is determined by $8I_n$ vectors of the root set $E_8$ [10].

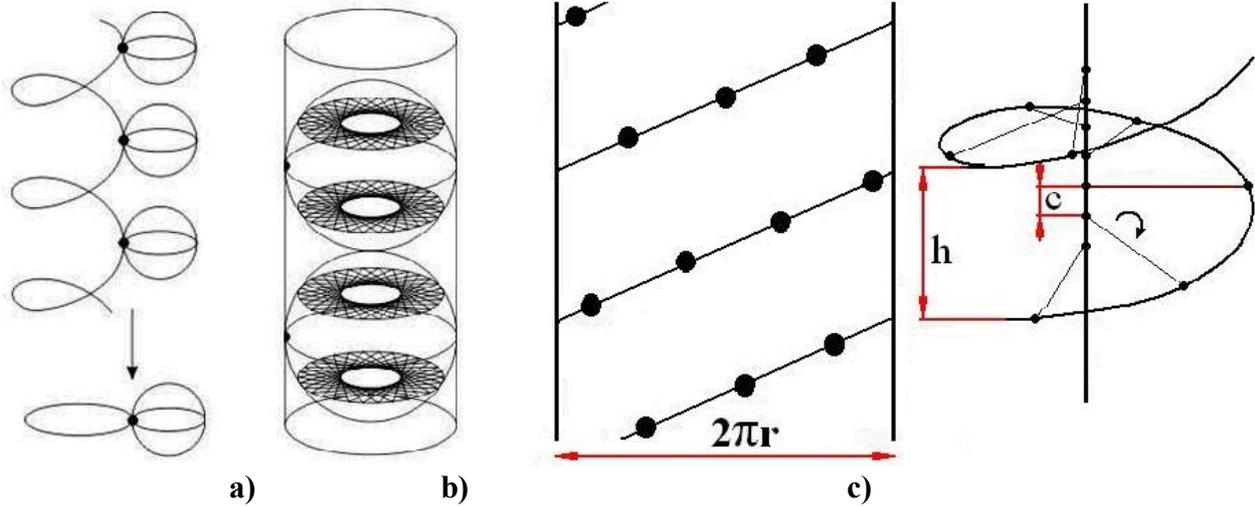

**a)                  b)                   c)**

*Fig. 2 a) Representation of a cover over a bouquet of the circle and the sphere in the form of spheres attached to a screw line. On every sphere $S^2$ shown is a solid common point (element) with the screw line, corresponding to a point (element) of a manifold on $S^1$.*
*b) Mapping of the vertices of an algebraic polytope to the spheres attached to the generate of a cylinder-like surface according to a.*
*c) A helix of pitch h and determined by the flat development of a cylinder of radius r. A rotation by the angle L/d determined by (5) and shift by c brings into coincidence two adjacent vertices of the helix.*

The difficulty is that no single closed subset of the n-dimensional Euclidean space is homeomorphic to a (n-1)-dimensional manifold. This difficulty can be circumvented by supplying components of the non-orientable surface with singularities. In particular, by defining every non-compact 2-dimensional manifold so that it is homeomorphic to some subset in $E^3$ (for example, it is possible to use normal non-compact form of the algebra su(2)). In the case of cell constructions, the necessary and sufficient condition to realize an oriented manifold $M^2$ (open or with an edge) as a plane domain, is that the index of intersection of any two 1-connected cycles is zero. The rationale for transition from a 2D surface and simplicial complexes defined on it to 3D cell structures and vice versa is the following: If the complex of a 3D manifold is characterized by $\pi_2(M) \neq 0$, then there is a 2D surface in it ($S^2$ or projective plane representing a non-trivial element $\pi_2(M) \neq 0$). Furthermore, if such a manifold M is oriented (for instance, $S^3$), then there is a $S^2$ in it, impressible in M.

A transition from the constructions given on $S^2$ to considering them on a plane (plane domain) is due to existence of a one-to-one relationship between their metrics – any metric-preserving transformation on $S^2$ is linear and orthogonal transformation of $E^2$. For the planar case there are only 4 root systems of rank 2: $A_1 \times A_1$, $A_2$, $B_2 = C_2$ (root lattice $Z_2$), $G_2$ (root lattice $A_2$) and $H_2$ (noncrystallographic system) [4], which may be used in classification of simple groups. Weil groups of Tits' system of rank 1, 2 are also Coxeter reflection groups, here the Weyl group of Tits' rank 1 with one generating element may be viewed as a dihedral group $D_m$ of order 2m, and of Tits' rank 2 as the group $D_\infty$. Note that $G_2$ cannot play the role of a root system for an irreducible system, as well as this group is a subgroup $D_4$ according to its decomposability, like $C_2$ is a subgroup of $A_3=D_3$. Then we obtain collections of roots not giving an infinite lattice but that can still be used to establish local lattice-like property. Correspondingly, the variants given are reduced to collections of 4 and 12 vectors (or 10), giving local lattice packings, satisfying the given conditions. Such approach will be demonstrated using an example of several polyhedral on $S^2$ and polytopes on $S^3$ in constructing the union $S^3 \to S^1 \cup S^2$. This allows one not only to determine singularities of manifolds on these spheres, but also regularities of common elements when forming a union.

There are several types of such mappings, in particular, for the direct product of spheres $M=S^i \times S^j$ and the bouquet of the "coordinate cross" $A=S^i \cup S^j=(S^i \times s_0'') \cup (s_0' \times S^j)$ contained in it, where $s_0' \in S^i$, $s_0'' \in S^j$ are marked points of spheres. In this case there exists a natural mapping

$$f: D^{i+j}=D^i \times D^j \to S^i \times S^j \qquad (4),$$

where f – is the direct product of two mappings of the form (of degree+1) $D^i \to S^j$ for $\partial D^i \to s_0'$ and $D^j \to S^j$ for $\partial D^j \to s_0''$ (in $S^i \times S^j$ a point $s_0=s_0' \times s_0''$ is defined). The mapping f transforms the boundary $\partial D^{i+j}=\partial(D^{i+j})=\partial D^i \times D^j \cup D^i \times \partial D^j$ into the manifold $A=S^i \cup S^j \subset M=S^i \times S^j$ so that $\partial D^i \to s_0'$ and $\partial D^j \to s_0''$. For i=1, j=2 have get $D^3=D^1 \times D^2$ (which corresponds to the handle) $H^3_1 \to S^1 \times S^2$ with the coordinate cross $A=S^1 \cup S^2$ attached in the point $s_0$, and the mapping $S^1 \cup S^2 \to X$, transforming the point $s_0$ into the point $x_0$, the initial point of the path. The manifold A mentioned above corresponds to a sphere cover without poles.

Thus, the problem of constructing a surface with given properties turns out to be related both to building a coordinate cross for $S^3$ (in order to then use a cover over the bouquet $S^1 \cup S^2$), as well as with constructing the plane torus $T^2$ as the disk $D^2$. Then more complicated constructions are built, determined by non-oriented nature of $S^2$ as well as using diverse variants of construction of cell simplicial complexes using handles and Mobius films. In particular, the functions mentioned above are used, possessing the property to be able to be represented as a direct sum of functions of each variable. Cell complexes allow one to graphically represent (see appendix B) a fiber bundle: E (fiber space)$\to$B(base) (F–fiber), because it is possible to write:

$$\sigma_E^{j+q} \cong \sigma_B^j \times \sigma_F^q \quad \text{and} \quad \partial \sigma_E^{j+q} \cong \sigma_B^0 \times (\partial \sigma_F^q) \quad (5),$$

if $\sigma_B^0$ – is the number of base vertices.

In order to determine locally-periodic properties when building cell complexes one uses mixed Abelian groups, containing torsionless subgroups (every element has finite periodicity) and torsion subgroups (using elements of unipotent subgroups, not equal no unity). Here the connecting role is played by the 1st and 2nd differential forms with fixed properties, which are given in various forms. To ensure that the conditions given above are satisfied, one uses a vector representation as well as the Gaussian property for a surface with locally vanishing curvature, given by an algebraic subsystem of vectors. Vectors that characterize positions of vertices of polyhedral on a helicoid surface, may be put into correspondence with elements of algebra $G_2$, which correspond to automorphisms of above lattices and are in correspondence with automorphisms of the lattice $E_8$.

Chevalley groups are generalizations (over local and commutative rings) of semisimple groups, and they are defined over Z, which allows one to use them to construct helicoidal locally periodic systems. Chevalley groups over K, by construction, are connected algebraic groups because they are generated by one-dimensional unipotent subgroups. If one uses [4] analogous constructions, replacing adjoint representations of algebras by other linear representations, in particular, projective ones, it is possible to obtain other types of similar simple groups. For example, quai-decomposable groups over finite fields lead to families of twisted finite groups. Over fields of characteristics 2 or 3 such twisted groups can be obtained by discarding the difference between short and long roots (which takes place in the systems $B_2$ or $G_2$).

A discrete version of mapping (2), (4), (5) is the polytope $\{q(2^n \bullet 24)\}$, n=0,1,2. In $E^3$ is possible in the form of its mapping into the polyhedron $\{2^n \bullet 24\}_q$ (рис.3. a, e), where each of 12 points $s_0$ maps into the center of the union of $2 \bullet 2^n$ vertices of the polyhedron. A shift of origin of coordinates into the deep hole of the lattice $E_8$ determines a sequence of coordination spheres of 16, 128, 448, 1024 … vectors. It allows one to single out a subset of 1152=128+1024 vectors of the second coordination sphere of $E_8$ that are correspondence with 240 reversible elements in Cayley's algebra, which has a group of automorphisms of type $G_2$. To realize a discrete variant (2),(4) the number q of elements on $S^1$ must equal 12. In the table [16] some regular and semi-regular polytopes, determined by subgroups of $F_4$, are given.

*TABLE[16] Regular and semi-regular polytopes were determined by subgroups of $F_4$.*

| Dynkin label | Cell counts by symmetry | | | | Element counts | | | |
| --- | --- | --- | --- | --- | --- | --- | --- | --- |
| | $W(SO(7))_L$ cells (24) | $(D_3 \times Z_2)_L$ cells (96) | $(D_3 \times Z_2)_R$ cells (96) | $W(SO(7))_R$ cells (24) | Cells | Faces | Edges | Vertices |
| (0001) | | | | 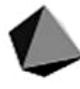 | 24 | 96 | 96 | 24 |
| (0010) | 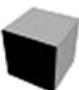 | | | 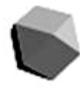 | 48 | 240 | 288 | 96 |
| (1011) | 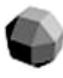 | 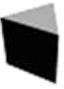 | 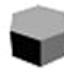 | 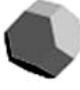 | 240 | 1104 | 1440 | 576 |
| (0111) | 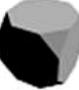 | 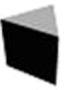 | | 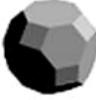 | 144 | 720 | 1152 | 576 |
| *(1111) | 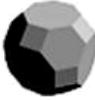 | 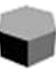 | 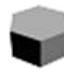 | 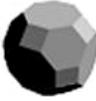 | 240 | 1392 | 2304 | 1152 |

* This orbit has the symmetry $Aut(F_4) \approx W(F_4) : Z_2$ of order 2304.

A each vertex of the polyhedron $\{2^n \bullet 24\}_q$ corresponds to a 12-element manifold $\{q_1\}$ on the sphere $S^1$ and the union $2\{q_1\}$ of two such helices forms a Q-chain, similar to <110> – a chain in diamond structure. The Q is mapped into the edge of $\{2^n \bullet 24\}_q$, which we shall denote by two-headed arrow (Fig.3.). For the polytopes $\{2^n \bullet 24\}_q$, for n=0,1,2, the point $s_0$ gets into the center of the union of $(2 \bullet 2^n)$ ends of arrows, namely, into the center of an arrow (fig.3.a), the center of a square (fig.3.d), and the center of the common edge of two 5-gons (fig.3.e.).

As is shown below, among the vertices of the polytope {576} (table) it is possible to select a 480-vertex submanifold. Analogous situation takes place with the polytope {1152} (table), where it is possible to select a 960-vertex submanifold. In fact, in both cases the problem is reduced to selection of various numbers of points (vectors), fixed under possible transformations of original manifolds (automorphisms of the system).

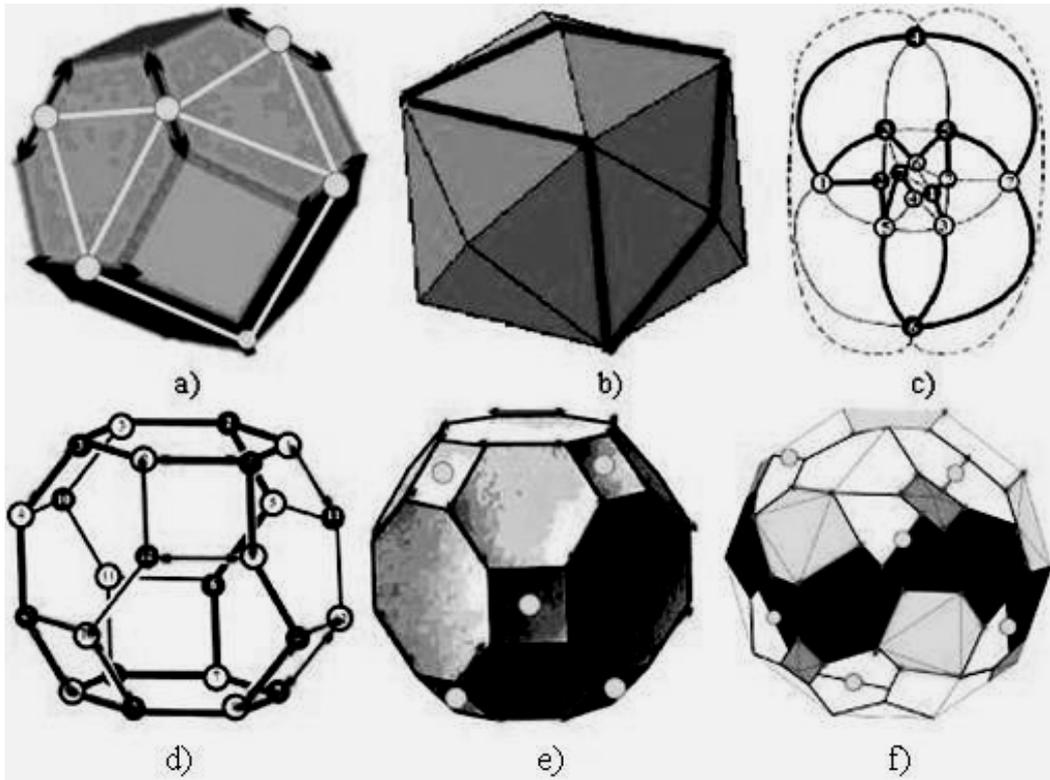

*Fig.3. a) 24-vertex truncated octahedron with 6 square and 8 hexagonal faces; the centers of 12 non-intersecting (red) edges-arrows are shown as yellow circles.*
*b). Partitioning a 14-vertex tetrahexahedron (dual to a truncated octahedron) into 6 hexacycles; edges of the tetra-hexahedron, marking one hexacycle, are shown in black.*
*c). Partitioning b) as a non-regular partitioning of the sphere, which can be embedded into the map $\{6,3\}_{2,1}$ on torus [9].*
*d). Truncated octahedron a), where a subsystem of 24 intersecting edges is selected.*
*e). A 48-vertex truncated cuboctahedron with 24 non-intersecting edges – arrows; the centers of 12 squares are shown as light dots. Vertices of squares are the ends of edges-arrows.*
*f). A 96-verrtex simple polyhedron with 5-, 6- and 7-gons and 48 non-intersecting edges-arrows. Alternating 5- and 7-gons, shown in light and dark shades of gray, form an equatorial belt; light dots show the centers of 12 edges, each of which is common for two 5-gons.*

### 3. The closed sequence of algebraic polytopes

The Chevalley group of type $G_2$ coincides (given a representation in the group $GL_7(K)$) with a torus and contains 3 non-unity involutions. Thus, in order to preserve, under the mapping (4), all the requirements considered above, the cut-out operation on a disc $D_0^2$ must determine the existence of three nearest vectors for any vector mapped from $E_8$ onto $S^2$. Correspondingly, the polyhedral $\{2^n \bullet 24\}$, n=0, 1, 2 must be simple (3 edges meet in every vertex), and their dual polyhedra be triangulated.

According to [15], an algebraic polytope may be diamond-like, if the polyhedron $\{2^n \bullet 24\}$ is simple and contains $2^n \bullet 36$ edges and $f_m$ m-vertex faces, m=4,5,6,7,8. According to Euler's theorem for simple polyhedra:

$$2f_4 + f_5 - f_7 - 2f_8 = 12 \quad \text{and} \quad f_4 + f_5 + f_6 + f_7 + f_8 = 2 + 2^n \bullet 12 \qquad (6).$$

The condition (6) is just necessary for the polytope to be diamond-like. All vertices of the polyhedron $\{2^n \bullet 24\}$ may be among the set of $2^n \bullet 12$ isolated edges, hence the sufficient condition for the polytope to be diamond-like is:

$$2^n \bullet 12 = V_Q \quad (7),$$

where $V_Q$ – is the set of Q-edges. In this case, in one vertex meet one Q–edge and two q-edges, into which are mapped the edges joining the vertices of Q–chains. The polyhedron, dual to the simple polyhedron $\{2^n \bullet 24\}$, n=0, 1, 2, determines a the specified number of vertex non-regular partition of the sphere into $2^n \bullet 24$ triangles, which allows one to give an appropriate simplicial (cell complex). For n=0 we have a 14-vertex partition $[4^6, 6^8]$ of the sphere by 36 edges into 24 triangles, where in each of 6 vertices 4 edges meet, and in each of 8 vertices 6 edges meet. Joining triangles by 4 into hexagons, we obtain a $\{6,3\}^3_{2,1}$-non-regular partition of the sphere into 6 hexagons (fig.3.b). This partition emerges out of the map $\{6,3\}_{2,1}$ on torus by discarding 3 edges forming a handle, turning the sphere into a torus. Uniting the fours of triangles of such partition into hexagons is possible in a right- or left-handed way (fig.3.b., d.), which selects a right- or left-handed triple of non-intersecting edges of the hexagon. Edges of the right-handed triple separate just hexagons, of the left-handed one – hexagons and squares. Mapping of the right triple into the left is determined, finally, by an involution from the group $E_8$. The maps $\{6,3\}_{n,1}$ represent incidence graphs of the finite projective planes PG(2,q). Therefore, taking into account all the mappings above allows one to obtain a sequence, connecting PG (2,2) (minimal of all finite projective planes) with the polytope $\{240\}$:

$$\{6,3\}_{2,1} \rightarrow \{6,3\}^3_{2,1} \rightarrow [4^6, 6^8] \rightarrow [4^6, 6^8]^* = \{2^0 \bullet 24\} \rightarrow \{2^0 \bullet 24\}_q \rightarrow \{10(2^0 \bullet 24)\} = \{240\} \quad (8),$$

where $\{6,3\}_{2,1}$ determines a regular (3 hexagons in every vertex) 14-vertex partition of torus into 7 hexagons, and $[4^6, 6^8]^*$ - dual to $[4^6, 6^8]$ (fig.3.b.) partition of the sphere, determining a truncated polyhedron (fig.3.a).

Doubling the number of vertices according to a certain law can be viewed as lifting a configurational degeneration. As a final result, such an approach gives gluing functions for the appropriate cover, allowing one to build a union of rods. If in hexagons of the truncated loaded octahedron $\{2^0 \bullet 24\}_q$, the right-handed triples of Q–edges are replaced by left-handed ones, one obtains a non-diamond-like loaded polyhedron $\{2^0 \bullet 24\}'_q$ (fig.3.g), for which the condition (7) is not met. This polytope may be viewed as doubly degenerate. For it, lifting the degeneration ensures (8) and transition to a loaded truncated cuboctahedron $\{2 \bullet 24\}_q$ (fig.3.d.). In analogy with the above, replacing in each hexagon $\{2 \bullet 24\}_q$ the right-handed triple of arrows for the left-handed one leads to a doubly degenerate non-diamond-like loaded polyhedron. Lifting the degeneration ensures transition to a 96-vertex diamond-like loaded polyhedron $\{2^2 \bullet 24\}_q$ with $2^2 \bullet 12$ isolated Q–edges, defined by the relations: $f_5 - f_7 = 12$ и $f_5 + f_6 + f_7 = 2 + 2^2 \bullet 12$ (fig.3.e.). Further replacement in each hexagon of the polyhedron $\{2^2 \bullet 24\}_q$ of the right-handed triple of arrows for a left-handed one does not lead to change in its type. Correspondingly, the subset of simple loaded polyhedra $\{2^n \bullet 24\}_q$, n=0, 1, 2 forms a closed sequence with respect to (determined by change of triples of Q–edges in 8 hexagons) the operation of degeneracy lifting for vertices.

By analogy (8), the map $\{6,3\}_{3,1}$ leads to a 26-vertex triangulation of the sphere $[4^{12}, 6^8, 8^6]$ and to a loaded polyhedron $\{2 \bullet 24\}_q$. The map $\{6,3\}_{5,0}$ (via the intermediate map $\{4,4\}_{5,5}$) leads to a 50-vertex triangulation of the sphere $[4^6, 5^{24}, 6^8, 7^{12}]$ and to the loaded polyhedron $\{2^2 \bullet 24\}_q$. Thus, the closed system of loaded polyhedra $\{2^n \bullet 24\}_q$, n=0,1,2 if determined by a sequence of map embeddings:

$$\{6,3\}_{2,1} \leftarrow \{6,3\}_{3,1} \leftarrow \{6,3\}_{5,0} \leftarrow \{6,3\} = A_2 \quad (9),$$

For which the universal cover is a hexagonal mosaic – lattice $A_2$. The map $\{6,3\}_{5,0}$ contains enough hexagons to determine all vectors of the root system $G_2$, given by the 2nd coordination sphere $A_2$ (fig.4.). Relations (8), (9) allow one to claim that there is a certain closed sequence of algebraic polytopes $\{q(2^n \bullet 24)\}$, n=0, 1, 2, beginning with the diamond-like polytope $\{240\}$ discovered by Coxeter [7,17].

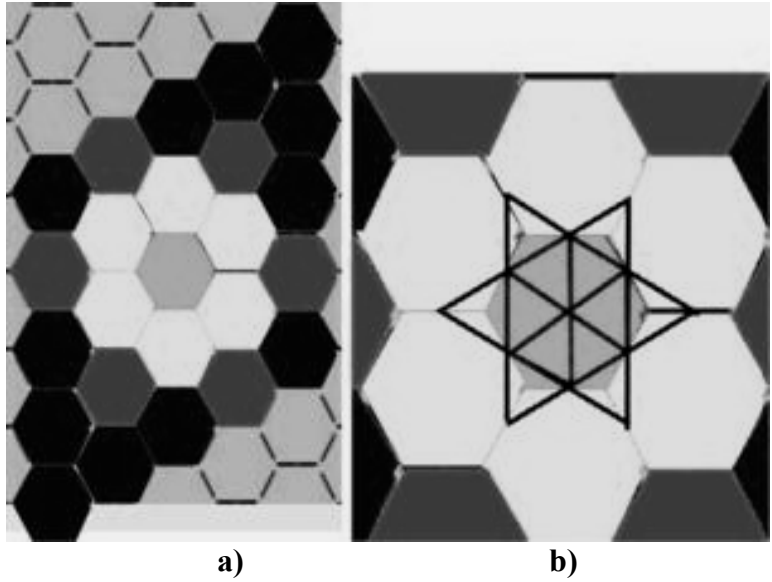

*Fig.4. a) The map $\{6,3\}_{5,0}$ of 25 hexagons of a hexagonal net, which contains 7 (gray and light gray) and 13 ( gray, light and light gray) hexagons determining the maps $\{6,3\}_{2,1}$ u $\{6,3\}_{3,1}$ b). The hexagon of the net a) and the union of 12 triangles of the lattice $A_2$, determined by vectors of the system $G_2$.*

A 3D rod substructure of a diamond-like algebraic polytope – the rod K corresponds to a face of the loaded polyhedron $\{2^n \bullet 24\}_q$ (fig. 3), formed by m edges, of which Q-edges are $m_Q$, for $0<m_Q<m/2$, m=5, 6, 7, 8. In $E^3$ the corresponding K channel is formed by congruent chains $2\{q_1\}$, (similar to <110> - diamond-like chain) divided into two types with respect to the channel. In a chain of the 1st type at every vertex the rod contains 3 edges (out of 4), and in a chain of 2nd type – 2 or 4 edges in alternating fashion. A chain of the 1st type is in correspondence with a Q-edge, of the 2nd type – with one vertex of Q-edge. According to [15], if the channel K is formed by chains of one type, it is an orbit of a screw axis, combining a rotation by the angle L/d and a shift by the vector h along the axis.

A point on $S^3$ in $E^4$ (with base vectors w,x,y,z) denoted by the quaternion $(z_1,z_2)$. Therefore the rotation symmetry of polytope denoted as $[R_{wx}(\alpha), R_{yz}(\beta)]$, where $R_{wx}(\alpha)$ signifies a rotation by angle $\alpha$ in the plane wx which leaves the plane yz invariant, $R_{yz}(\beta)$- a rotation by angle $\beta$ in the plane yz which leaves the plane wx invariant [17]. In fact, the polytope {240}, initiating a closed sequence of polytopes, is a union of 4 chains 30/11, in the interstices between which are 6 channels 40/9 (fig. 3.a). The generating relations are $(30/11)^3=(40/9)^4=(10)^1$ for rotations in the plane wx, determined by generating relations for the group $(O'\times Y')/2$ of the polytope [17, 18]. The polytopes following the {240} polytope in the sequence, determined by the polyhedron $\{2\bullet 24\}_q$ (fig. 3.d), is characterized by generating relations

$$(30/11)^3=(40/11)^4=(40/9)^4=(10)^1 \qquad (10),$$

that determine the appropriate channels (fig. 5). The minus sign determines different chirality of axes. Thus, a topologically stable helicoid-like structure $\Omega$, determined by an algebraic polytope must satisfy the following relations:

$$\Omega \leftarrow <L/d|\lambda\cdot h>2\{q\}_k \leftarrow \{q(2^n\bullet 24)\} \leftarrow E_8 \qquad (11)$$

where the rotation of L/d is given by (5), $\lambda$ is an integer; k=1 or 2; h=2.4r, r - the radius of this cylindrical - like surface, on which can be displayed $\Omega$. In common case, the radius-vectors **r**(u,v) are used, where v corresponds to the angle φ in the cylindrical system of coordinates is determined in $E^3$ by vector $\mathbf{r}(u,v,\alpha)=\cos\alpha\ \mathbf{r_1}(u,v)+\sin\alpha\ \mathbf{r_2}(u,v)$, where the isothermal coordinates (u,v) is parameters of a representation after a period and $\alpha \in [0,\pi/2]$; for $\mathbf{r}(\alpha=\pi/2)$ corresponds to helicoid, and $\mathbf{r}(\alpha=0)$ corresponds to catenoid (spanning more detail in part II). By L/d=n, n=1,2,3,4,6 and $\lambda \cdot h=t/n$, where t belongs to lattice $D_3$ ($D_3 \subset D_4$) or $A_2 \times A_1$ ($G_2$ is determined by second coordination sphere of $A_2$), we obtain the symmetry operator of space group.

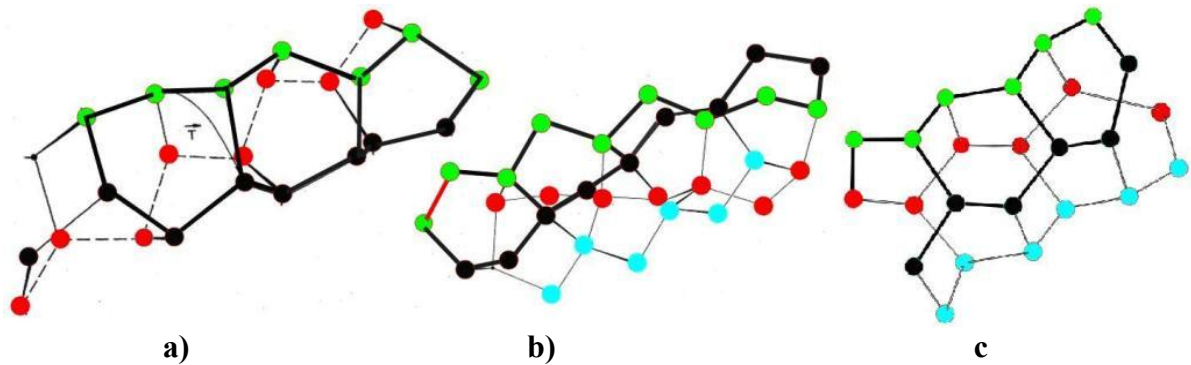

**a)** **b)** **c**

*Fig.5. a, b, c. Channels 30/11, 40/9 [17, 18] and 40/11, as the unions как of <110>-diamond-like chains (green ones, red ones, black and blue ones), which correspond to hexagonal, square, and octagonal faces of the polyhedron fig.3.e. The union of the green and red chains is emphasized by solid black lines.*

## 4. Topological stability and bifurcation points of discrete helical - like substructures determined by the algebraic polytopes

All non-planar non-congruent complete ruled (when using polyhedral constructions this factor is essential) minimal surfaces may be realized as a 1-parameter family of helicoids with the parameter h – screw pitch [3]. A catenoid determined by one such surface is locally isomorphic to a helicoid belonging to a class of minimal surfaces allowing for conformal coordinates. Helicoid may be represented as an infinite – valued surface over a catenoid (folds remain within the class of minimal surfaces) or onto a sphere without poles [3]. Correspondingly, if we have an axis formed by two circumferences sharing an axis (they are of the same radius and are situated in parallel planes), then in order to describe all minimal surfaces closing in on the contour it suffices to describe all catenoids closing in on it. A complete solution of this problem was given already by Poisson. Denote by h the distance between the circumferences (of radius r). Then it turns out that for small h a solution gives two catenoids, one of which is close to a cylinder (stable configuration), and the other to a cone (unstable configuration). A general approach will be considered in the next section, while for some helicoidally similar structures the parameters of a bifurcation point may be determined in an approximation of their cylindrically similar surface.

Drawing in cylindrical coordinates a graph describing such a catenoid as a function of h, the corresponding solution (related to the solution of the equation of the form cth$x=x$) [3]) for h=$h_{кр}$, leads to the relation $h_{кр}$=2, 4·r, where r is the cylinder radius corresponding to the enclosing film. For some value h=$h_{кр}$ both catenoids form a single configuration, covering each circumference by a plane disc. Thus, the equation mentioned gives a bifurcation (instability) point of the catenoid upon changing the distance between the circumferences of the catenoid, when the latter decomposes into a stable cylinder and an unstable cone.

Recall [1,vol.3, 19], that in a bifurcation point (non-degenerate for a Morse function -see appendix C)) topological regularity is broken; at the same time a cell structure considered in the preceding section must appear on the manifold. Building bundles giving minimal surfaces described by Weierstrasse representation and an m-valued surface covering the catenoid, corresponds to a diffeomorphism in agreement with a Gaussian map. In our further treatment we employ, in fact, a technique of cutting and gluing of polyhedra, used in algebraic topology [1].

The stereographic projection of $S^3$ on $E^3$ is a union of toroid surfaces. Thus, a transition (2), (4) from $S^3$ to a universal cover may be viewed as corresponding to a Gaussian map of $S^3$ into a tubular surface in $E^3$ [1]. Giving a polytope on $S^3$ corresponds to a discrete variant of (2), realized upon introduction of an exterior metrics, when for any two points of helicoid's surface the distance between them equals Euclidean distance in $E^3$. In such case one may define in $E^3$ a union of locally cylindrical (tubular) similar surfaces by the screw curve corresponding to $S^1$ from (2). Then only some points of this line (namely, the vertices of the cell complex) lie on the said surface, determined by the relation (5). The curvature (k) and torsion ($\chi$) of a spatial curve in Euclidean space form a complete set of its geometric invariants. Therefore, if a curve is given by vectors (for example, by an orthonormal basis for the speed along the curve of normal and binormal vectors [3]), then, under certain conditions, the latter may be given with values for a basis in appropriate subalgebra. For the relation k=c$\chi$ to hold, where c is constant, between these parameters it is necessary that there exists a vector u, for which <uv>=const. Then we define a Darboux' vector (in correspondence with [3]), as well as other vectors as elements of the algebra of the group $G_2$, not contained in the original root basis. The latter implies a transition to various groups (subalgebras) of Shevalley $G_2$, which allow one to construct a local lattice system. In the end, the homogeneity of a screw line is given in the sense that the distance between turns (on a cylinder) is related in certain way with the length of the turn. In the discrete variant this means that the number (not necessarily integral) of vertices per turn is constant.

Let S be some surface given by Weierstrass representation (U,w, (aw+d)$^m$)) ,a,b$\in$**C**$\neq$0, where m is a non-zero integer, U$\subset$**C** is some subdomain of the complex plane (as a result of complexification for discrete constructions). The surface S is characterized by zero index if the image of the domain U under a Gaussian map is contained in some open subset K of the sphere $S^2$. The subset K can be defined either as an open hemisphere of the sphere $S^2$ without the pole, or as the part of the sphere $S^2$ without one pole. It contained between two parallel planes separated from the center (zero) of the sphere by the distance th $t_0$, where $t_0$ – is the only root of the equation

$$\text{cth } t_0 = t_0 \quad (12)$$

A local representation of minimal surface is reducible to using solutions of the form $\varphi=\partial r/z$, which corresponds to a holomorphic radius vector for $\varphi^2=0$ (for regularity of the surface in conformal coordinates, which is determined by giving nilpotent groups and, in particular, p-groups and corresponding p-algebras). The Gaussian map itself allows to conserve angles between vectors of the vector manifold under diffeomorphisms (tangent mapping). The Weierstrass' representations allow one to give a catenoid as well as a complete helicoid. In the general case, an associated family for some minimal surface M (for instance, helicoid's or catenoid's) consists of locally isometric minimal surfaces (incongruent pairwise, as a rule).Thus, the problem of constructing a surface with given properties turns out to be related to both building a coordinate cross for $S^3$ (in order to then use a cover over the bouquet $S^1 \cup S^2$), using a vector 1-form and polytopes, and then realizing via cell complexes of the disks and the plane torus $T^2$ (as a disk $D^2$). A transition to considering locally-periodic structures, as well as a local approach to consideration of the entire complex of problems, is to a considerable degree related to the fact that the indices of periodic minimal surfaces (including Riemann-Shwartz and Sherk surfaces) are infinite.

Consideration of the disks $D^2$ leads to using constructions like the root system $G_2$, which consists of vectors of the 1$^{st}$ and 2$^{nd}$ coordination spheres of the lattice $A_2$. In fact, the algebra $G_2$ may be viewed as the group of automorphisms of a 8D Cayley's algebra, which generates a quaternion subalgebra and is in correspondence with the Hamming's code $H_8$. The whole Cayley's numbers (octonions) are identified with $E_8$ upon identifying the 240 invertible elements of norm 1 of Cayley's algebra with the 240 minimal vectors (the 1$^{st}$ coordination sphere) of the $E_8$ lattice. By quasi-reducibility $G_2$ corresponds to $D_4$, which allows one to use Borel subgroups. The choice of $G_2$ is also related to the necessity of using a non-diamond-like polytope {240} which allows one to construct a diamond-like polytope {480}, which is a union of three polytopes {160}. In such case, uniting three helices into a super-helix corresponds to the polytope {480}.

Describing a closed minimal net (which in the problem under consideration appears as a family of edges) on a 2D oriented surface (generalized Steiner's problem) is simplified using a torus $T^2$, which can be defined as a parallelogram with identified opposite sides. Then it is possible to build a natural map $\pi: E^2 \to T^2$ (universal cover), by putting into correspondence with every point on the plane a point on the torus so that the set of parallelogram's vertices generates a lattice (sub-algebra). In order to describe isometry classes of such tori it is possible to use classes of lattices, because their similarity ensures the similarity of the appropriate classes of tori. Furthermore, it turns out that they class of such tori is in one-to-one correspondence with the set of lattices generated by only certain types of lattices. Correspondingly, it is possible to move from considering one type of closed minimal nets toward other non-equivalent minimal nets, where small perturbations of the plane torus $T^2$ (in the class of plane tori) does not destroy closed minimal net.

Recall that this equation also gives a bifurcation (instability) point [19] of the catenoid on changing the distance between the circumferences of the catenoid, when the latter decomposes into a stable cylinder and an unstable cone. Because diffeomorphisms are used in the construction of fibrations, it is important to note that minimal surfaces, described by Weierstrass representation and m-valued face of the catenoid, correspond to a diffeomorphism in agreement with Gaussian map. In part 2 is it planned to build more complicated constructions, determined both by non-oriented nature of $S^2$ (doubling), as well as using different variants of construction of simplicial cell complexes. In particular, using handles and Mobius films to build constructions using the functions pointed out above, which possess the property of being represented as a direct sum of functions of each variable.

## 5. Conclusion

In the general case a discrete system consisting of subsystems is preferable as compared with a continuous one because it possesses greater stability. In fact, in such a system every subsystem may exist in a small number of possible states, because it, as a rule, ignores small displacements from the normal values of various physical parameters of the system, recovering (in the original form) one of its allowed states. In a continuous system small perturbations accumulate, which can lead to essential changes in the system itself. Thus, in order to obtain physically meaningful results one has to use the solutions corresponding to finite stationary Hamiltonian systems, for which the system of levels is also discrete.

It is shown in the present work that in order to achieve this it is necessary to use 3D lattice constructions, corresponding to local invariant transformations. As is known, the problem of classification of 3D manifolds is not only not solved, but it is also not known whether it is algorithmically possible to solve it [1, vol.3,]. However, if one limits himself to consideration of diffeomorphisms (or automorphisms, which in fact is used when considering topological structural elements) of the surface enclosing the volume of the 3D structure in question, then it has turned out to be possible to separate certain classes of 3D manifolds. The latter has determined the necessity of considering the surfaces, related to minimal ones (locally minimal ones) and containing singular points that it was necessary to put into correspondence with selected lattice type as well as automorphisms of the system.

It has been established in this work that the polytope {240}, defined by Coxeter, is in the beginning of the closed system of algebraic polytopes: $\{q(2^n \bullet 24)\}$, n=0,1,2, where q=10 or q=12. The relationships between polytopes of this sequence with fiber bundles (covers) and cell complexes have allowed us to give a definition of non-crystallographic helicoid-like substructures. The topological stability of such substructures is determined by their relatedness to Weierstrass' representation, as well as the condition that the instability index of the surface equals zero. In our next paper (which is going to be the continuation of the present one) it will be shown that such *a*

*priori* derived special local lattice packings of cell complexes enclosed by a helicoidally similar surface, are realized in the form of the α-helix and various forms of DNA structures.